\documentclass[preprint]{aastex62}

\received{April 20, 2018}
\accepted{May 22, 2018}
\submitjournal{ApJ}

\shorttitle{Two runaway stars as unidentified {\it Fermi} sources}
\shortauthors{E. S\'anchez-Ayaso et al.}

\begin{document}

\title{Possible association of two stellar bowshocks with unidentified {\it Fermi} sources}

\correspondingauthor{E. S\'anchez-Ayaso}
\email{esayaso@ujaen.es}

\author[0000-0002-3746-0612]{E. S\'anchez-Ayaso }
\affiliation{Departamento de F\'isica (EPS),\\Universidad de Ja\'en, \\Campus Las Lagunillas s/n Ed. A3 Ja\'en,\\Spain, 23071}

\author{Mar\'{\i}a V. del Valle}
\affiliation{Institute of Physics and Astronomy, \\University of Potsdam, \\14476 Potsdam, Germany}

\author{J. Mart\'{\i}}
\affil{Departamento de F\'isica (EPS),\\Universidad de Ja\'en, \\Campus Las Lagunillas s/n Ed. A3 Ja\'en,\\Spain, 23071}

\author{G.E. Romero}
\affiliation{Instituto Argentino de Radioastronom\'{\i}a (CONICET/CICPBA), \\C.C.5, (1894) Villa Elisa, Buenos Aires, \\Argentina\\}
\affiliation{Facultad de Ciencias Astron\'omicas y Geof\'{\i}sicas, \\Universidad Nacional de La Plata, \\Paseo del Bosque, B1900FWA La Plata, \\Argentina}

\author{P.L. Luque-Escamilla}
\affiliation{Departamento de Ingenier\'{\i}a Mec\'anica y Minera, \\Escuela Polit\'ecnica Superior, Universidad de Ja\'en, \\Campus Las Lagunillas s/n, A3, \\23071 Ja\'en (Spain).\\}

\begin{abstract}

   The bowshocks of runaway stars had been theoretically proposed as gamma-ray sources. However, this hypothesis has not been confirmed by observations up to date. In this paper, we present two runaway stars ($\lambda$ Cep and LS 2355) whose bowshocks are coincident with the unidentified \textit{Fermi} gamma-ray sources 3FLG J2210.1+5925 and 3FGL J1128.7-6232, respectively. After performing a cross-correlation between different catalogues at distinct wavelengths, we found that these bowshocks are the most peculiar objects in the {\it Fermi} position ellipses. Then, we computed the inverse Compton emission and fitted the {\it Fermi} data in order to test the viability of both runaway stars as potential counterparts of the two high-energy sources. We obtained very reasonable values for the fitted parameters of both stars. We also evaluated the possibility for the source 3FGL J1128.7-6232, which is positionally coincident with an HII region, to be the result of background cosmic-rays protons interacting with the matter of the cloud, as well as the probability of a pure chance association. We conclude that the gamma rays from these {\it Fermi} sources might be produced in the bowshocks of the considered runaway stars. In such a case, these would be the first sources of this class ever detected at gamma rays.

\end{abstract}

\keywords{gamma-rays: ISM --
             radiation mechanisms: non-thermal --
              stars: winds, outflows}

\section{Introduction} \label{sec:intro}

There is an important significant number of unidentified sources of high and very high-energy photons. Even with the improved capabilities of the {\it Fermi} Large Area Telescope (LAT)  
\cite[see][]{2009ApJ...697.1071A}, the latest release of its 3FGL catalogue \citep{2015ApJS..218...23A} contains 1059 GeV sources without a counterpart at lower energies. This represents about 35\% of the total.
The situation is not much better at very high-energies where the latest edition of the TeVCat catalogue \citep{2008ICRC....3.1341W} includes 27\%  of unidentified sources. Many of these unassociated sources are located in our Galaxy. However, the search for counterparts among well-known gamma-ray emitters in the Milky Way (objects such as pulsars, supernova remnants, molecular clouds or compact binaries among others) is not always successful.  

Among the more exotic gamma-ray emitters proposed in the literature we can find the bowshocks of runaway massive stars. \cite{2012A&A...543A..56D,2014A&A...563A..96D} provided theoretical arguments for these objects to be high-energy sources.  In these models the bowshocks of luminous runaway stars act as the acceleration sites of relativistic particles. Gamma rays result from inverse Compton (IC) interactions of relativistic electrons with background thermal photons and, to lesser extent, from $pp$ collisions. The expected luminosities are, relatively modest in the GeV domain ($\sim 10^{31}$-$10^{33}$ erg s$^{-1}$) and the detection of such bowshocks in gamma rays is only plausible if they are nearby. So far, only two candidates have been proposed: $\zeta$ Ophiuchi, which might be detected by the ground-based detector Cherenkov Telescope Array (CTA) in the future as well as by the {\it Fermi} satellite, and HD 195592, which was proposed to be associated with the {\it Fermi} gamma-ray source 2FGL J2030.7+4417 \citep[see ][respectively]{2012A&A...543A..56D, 2013A&A...550A.112D}. This latter source, however, was recently claimed to be associated with a pulsar. Therefore, there is currently no bowshock clearly associated with an unidentified {\it Fermi} source.

At radio, non-thermal radiation has been found in the bowshock of BD+43$^{\circ}$3654 with the Very Large Array (VLA) \citep[see][]{2010A&A...517L..10B}. This indicates the presence of relativistic particles and motivates further efforts in the search of high-energy counterparts.

\citet{2014A&A...565A..95S} performed the first systematic search in the MeV and GeV emission from the 27 bowshocks listed in the E-BOSS catalogue \citep{2012A&A...538A.108P}. They analyzed 57 months of {\it Fermi}-LAT data, and found no significant detection. They reported upper limits for all  sources, ruling out the most optimistic model predictions for
$\zeta$ Ophiuchi by a factor of 5. In addition, the High Energy Stereoscopic System (H. E. S. S.) Collaboration \citep{2017arXiv170502263H} performed a similar analysis of 32 sources listed in the second E-BOSS catalogue \citep{2015A&A...578A..45P}.  None of these sources showed detectable emission at $E$ $>$100 GeV. 
Despite the efforts made to find high-energy radiation from the bowshocks  of massive runaway stars, a reliable detection still remains elusive.

Similar searches were carried out in X-rays using data from XMM-{\it Newton} \citep{2016ApJ...821...79T,2017ApJ...838L..19T,2017MNRAS.471.4452D}, with no positive identification. On the basis of energy considerations for all sources in the E-BOSS catalogue, \citet{2017MNRAS.471.4452D} anticipate that the detection of inverse Compton emission from the bowshock of massive runaways with current X-ray facilities is very unlikely.

In this work, we report two runaway stars whose bowshocks are firmly candidates for being responsible of the gamma-ray emission in two respective unidentified \textit{Fermi} sources. One of them is a massive runaway that was not previously known and is reported here by first time.

\section{Search for possible counterparts of bowshock nature to unidentified Fermi sources}

In a first step, we cross-matched the luminous star catalogues by \cite{1959LS....C01....0H}  and \cite{1971PW&SO...1a...1S}  
with unidentified gamma-ray sources from the latest release of the {\it Fermi} 3FGL catalogue. 
This was done for low galactic latitudes $|b|$$<$5$^{\circ}$ because we are interested on galactic sources in the Milky Way.
In a second step, positional matches were further investigated by examining the presence of arc-shaped bowshocks that could trace the possible runaway nature of the stars.
This kind of structures are easily revealed by the far infrared (IR) emission from the heated dust accumulated as the star sweeps the interstellar medium (ISM).
This cross-check was often facilitated by the different releases of the E-BOSS catalogue in \cite{2015A&A...578A..45P} and references therein.

As a result, here we found two new promising candidates: the stars  $\lambda$ Cep (HD210839, HIP109556) and  LS 2355 (HD99897, HIP56021). The physical parameters of the stars are shown in Table \ref{parametersLamCep} and the details about their radial velocity, proper motions and distances are given in Appendix A. We tentatively associate them with the
unidentified {\it Fermi} sources 3FLG J2210.1+5925 and 3FGL J1128.7-6232, respectively, as no other remarkable object appears inside their error ellipses. While $\lambda$ Cep is part of the E-BOSS catalogue, it is worthwhile to note that the star LS2355 represents a new addition to the known population of stellar bowshocks.
In the following sections we describe the observational parameters of both objects. This information is used as an input for apply  a simple model  in order to fit the observed  spectral energy distribution (SED) \citep[following][]{2017MNRAS.471.4452D}, and then to assess the possibility of a physical association with the {\it Fermi} high-energy emission.

\section{$\lambda$ Cep as a possible gamma-ray source}

\begin{figure}
\centering
\includegraphics[width=8.5cm]{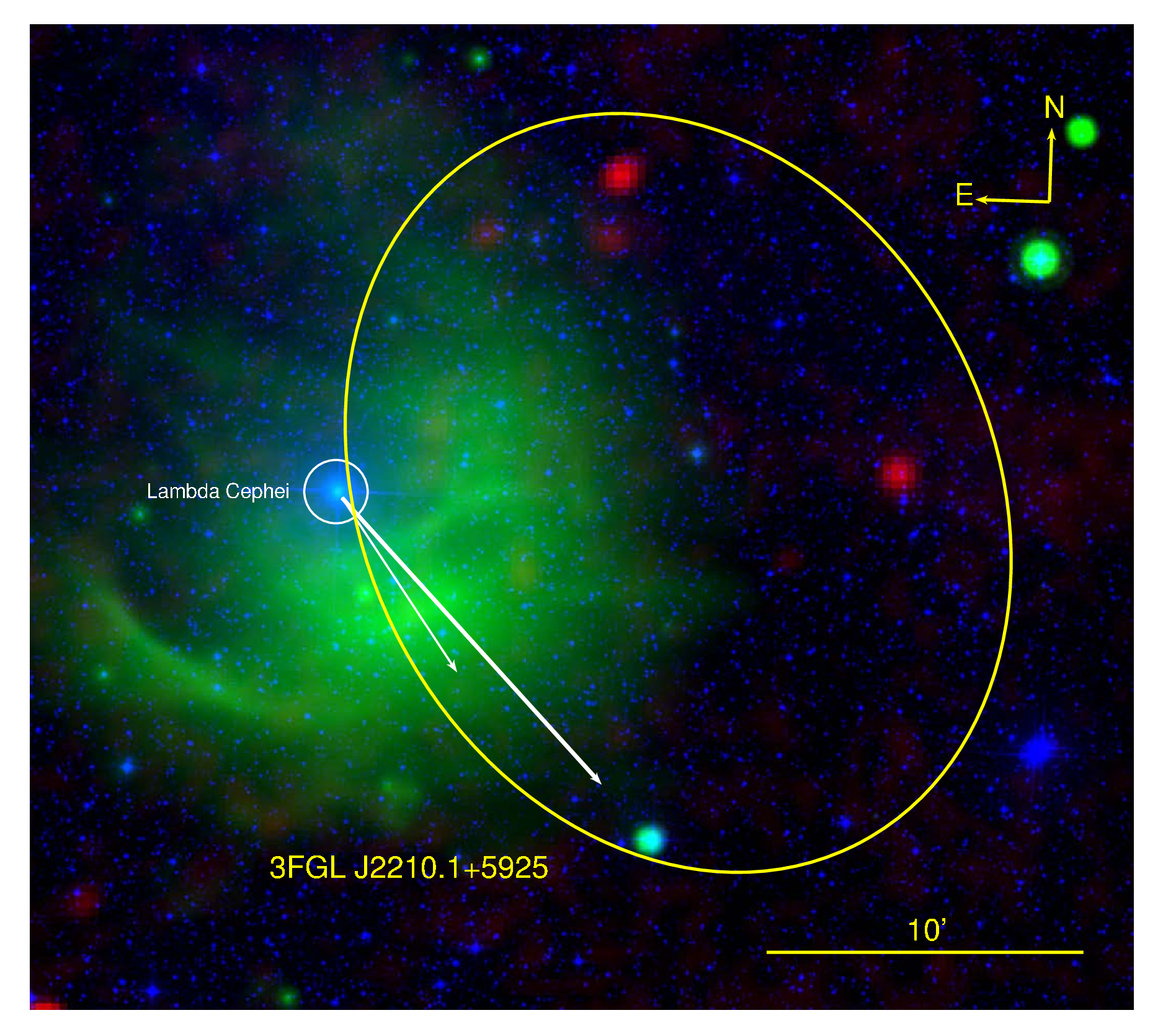}
\caption{View of the 95\% confidence ellipse of the unidentified gamma ray source 3FGL J2210.1+5925 projected in the direction of the shock front created by the runaway star 
$\lambda$ Cep. The layers of this  trichromatic composition shows the NVSS radio emission at 1.4 GHz in red, the {\it WISE} infrared emission at 22 $\mu$m in green and the optical emission of the DSS2 red plate in blue. The white circle marks the position of the runaway star. The bowshock is the most peculiar object of the field. Both the Galactic rotation uncorrected (thin) and corrected (thick) proper motion vectors are also plotted.}
 \label{elipse_Lam}
 \end{figure}


The star  $\lambda$ Cep, with spectral type O6.5I(n)fp \citep{2011ApJS..193...24S},
 is one of the most massive runaway stars known up to date. It is located at a distance of about 0.6 kpc according to the parallax provided by the {\it Hipparcos} satellite  \citep*[see  also Appendix \ref{pmdis}]{2007A&A...474..653V}. Unfortunately, this star is not yet included the latest {\it Gaia} data release.

Figure~\ref{elipse_Lam} shows the 95\% confidence error ellipse of the unidentified gamma ray source 3FGL J2210.1+5925 over a trichromatic composition that shows 
the radio, IR and optical appearance of the field as red, green and blue layers, respectively. The composition displays the corresponding image at 1.4 GHz
from the NRAO VLA Sky Survey \citep[NVSS,][]{1998AJ....115.1693C} in red, the Wide-field Infrared Survey Explorer (WISE) emission at 22 $\mu$m in green and the Digitized Sky Survey optical emission of the red plate in blue \citep{2003MNRAS.342.1117M, 2010AJ....140.1868W, 1990AJ.....99.2019L}. The bowshock around $\lambda$ Cep appears as the most peculiar object compatible with the {\it Fermi} ellipse position (see Fig.~\ref{elipse_Lam}).
Historically, this extended feature around $\lambda$ Cep was first detected in  {\it IRAS} images \citep[see][]{1988ApJ...329L..93V}. 
The volume of the paraboloid limited by the star and the  bowshock vertex amounts to  $(1.3 \pm 0.5) \times 10^{56}$  cm$^3$. The estimated velocity relative to the star Regional Standard of Rest (RSR)
 gives $70 \pm 6$  km s$^{-1}$ (see Appendix \ref{rotgal}). After applying the correction for Galactic rotation, the position angle of the RSR velocity changes by a few degrees improving the expected alignment with the bowshock vertex. From the image we can estimate the distance $R$ from the star to the midpoint of the bowshock. Correcting from the angle between the peculiar RSR velocity and the observer, estimated to be $\sim$ 30 $\pm$ 5\,$^{\circ}$, we obtain $R \sim 1.6$\,pc.

 This star has a high bolometric luminosity and a powerful wind. The wind properties quoted in Table \ref{parametersLamCep}  have been taken from \citet{2007A&A...473..603M}. They combine to provide a mechanical power $P_{\rm w} = 0.5 \dot{M} V_{\rm w}^{2} \sim 10^{37}$\,erg\,s$^{-1}$.


\begin{table}
\label{parametersLamCep}      
\centering     
\caption{Physical parameters for the stars}
\renewcommand{\arraystretch}{0.5}
\renewcommand{\tabcolsep}{0.06cm}
\tiny{
\begin{tabular}{c c c c}          
\multicolumn{2}{c}{Parameter} & $\lambda$ Cep & LS2355 \\    

\hline\hline  
&&&\\

SpT& Spectral type
&  O6.5I(n)fp
&O6.5IV\\    

$V_{\rm w}$ [$$\,km\,s$^{-1}$]& Wind velocity
& $2.25 \times 10^{3}$& $2 \times 10^{3}$\\

$\dot{M}$ [\,$M_{\odot}$\,yr$^{-1}$]& Wind mass-loss rate
& $6.8 \times 10^{-6}$ & $10^{-6}$\\

$P_{\rm w}$[$$\,erg\,s$^{-1}$]& Mechanical wind power
& $\sim 10^{37}$ &  $\sim 10^{36}$\\

$R$ [cm]& Stagnation point
& $5.06 \times 10^{18}$  & $1.79 \times 10^{18}$\\

$  \chi \sim P_{\rm w}/P_{\rm e}$ & Fraction of the wind power
& 0.6 & 2.7 \\

$L_{\rm *}$ [$$\,erg\,s$^{-1}$]& Bolometric luminosity
& $\sim 2 \times 10^{39}$ & $\sim 2 \times 10^{39}$\\

$B$ $[\mu G]$ & Minimum magnetic field
 & 0.06  & 0.1
\\   

\hline                                            

\end{tabular}
}
\end{table}


\subsection{Spectral energy distribution of the $\lambda$ Cep bowshock}


\begin{figure}
\centering
\includegraphics[width=9.5cm,angle=0]{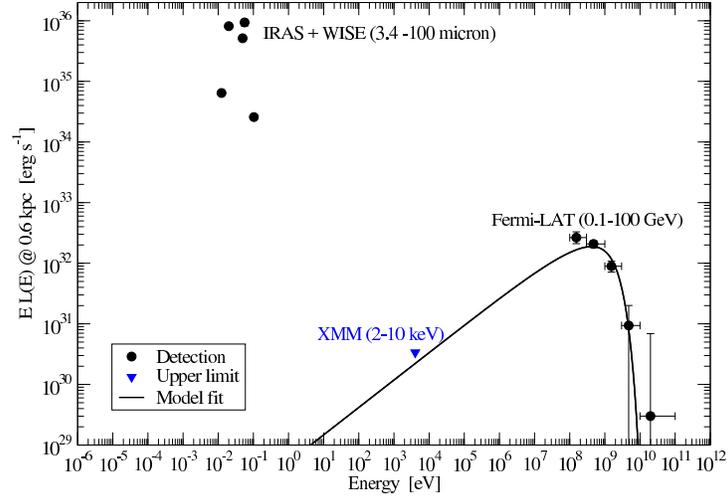}
\caption{Fitted SED for $\lambda$ Cep bowshock at an assumed distance of 0.6 kpc. 
{\it IRAS} (12-100 $\mu$m),  {\it WISE} (22 $\mu$m), X-ray upper limit (2-10 keV) and {\it Fermi}-LAT data are shown. }
  \label{sed_lam}
  \end{figure}

In Figure \ref{sed_lam}, we display the SED of the $\lambda$ Cep bowshock together with the {\it Fermi}-LAT spectrum
of 3FLG J2210.1+5925. Axes are logarithmically distributed. The energy 
 $E$ is in eV, and the vertical axis shows the monochromatic luminosity times the energy, $E L(E)$ in erg s$^{-1}$.
In addition to the IR flux density, taken from {\it IRAS} and completed by performing aperture photometry on the {\it WISE} 22 $\mu m$ image,
an upper limit to the extended X-ray emission is also included. 

Given the absence of diffuse X-ray radiation associated with the bowshock,
we proceeded  to calculate a 4$\sigma$ upper limit from XMM-{\it Newton} archival data (Obs. ID 0720090501).
A power-law spectrum with $\Gamma = 2$ and a hydrogen column density of $N_H \sim 5.3 \times 10^{22}$ cm$^{-2}$ has been assumed for the
fitting. This value is consistent with the observed color excess for $\lambda$ Cep $E(B-V) \simeq 0.56$ mag, equivalent to an interstellar extinction of
$A_V \simeq 5.3 \times 10^{-22}  N_H \simeq 1.7$ mag.
Concerning radio emission, we preferred not to include an upper limit in this SED plot because it could not be physically meaningful given the
lack of sensitivity of the NVSS data to angular scales as large as those of the bowshock ($\sim20^{\prime}$).
Numerical values for the SED data points are given in the Appendix (Table \ref{fluxesLam}).

Following the approach by \citet{2017MNRAS.471.4452D}, we use a simple model to fit the observed gamma emision with the aim of establishing whether it is physically possible for the bowshock of  $\lambda$ Cep to generate the \textit{Fermi} source.

\subsection{Model and fitting}\label{sec:model}

We assume that a population of relativistic electrons is accelerated and injected in the system; the injection function is a power law in energy of index $\alpha$, as expected in diffusive shock acceleration (DSA), and normalized to the power in relativistic electrons. We adopt here a one-zone model where the electrons and the target fields are homogeneous and isotropic, localized in the same region. The particles are most likely accelerated at the reverse shock in the wind, that is adiabatic and strong with $V_{\rm s} \sim V_{\rm w}$; therefore the power in relativistic electrons $P_{\rm e}$ is a fraction $\chi$ of the wind power $P_{\rm w}$. As we can see from Figure\,\ref{elipse_Lam} the {\it Fermi} ellipse does not include the whole bowshock: only about $\sim$ 60\% of its area\footnote{If the actual area of the bowshock enclosed by the {\it Fermi} ellipse is between  10\% and 100\%, then adopting a value of 60\% gives an error of less than a factor of 6, in which case it would be absorbed by $\chi$.} is enclosed. Consequently, we adopt  $P_{\rm e} \sim \chi (0.6 P_{\rm w})$. 

The relativistic electrons loss energy mainly by synchrotron and IC radiation. The first process is caused by the interaction with the local magnetic field $B$, while the second is produced through collisions with IR photons\footnote{The photon radiation field of the Cosmic Microwave Background, with $E_{\rm ph} \sim 1.9\times 10^{-4}$\,eV, might also be important for IC scattering; however the energy density of this field, $\sim 1.2\times10^{-13}$\,erg\,cm$^{-3}$, is much smaller than the local field produced by the IR photons. The latter can be estimated as $L_{\rm IR}/(c\,A_{\rm bs}) \sim 10^{-11}$\,erg\,cm$^{-3}$, with $A_{\rm bs}$ the area of the bowshock.}. The particles also suffer from transport effects being convected away by the shocked stellar wind. The local hadronic contribution is negligible since protons are removed by convection long before they can cool \citep[see e.g.,][]{2012A&A...543A..56D}.

The IR luminosity is emitted by the dust accumulated in the bowshock and heated by the stellar photons. This emission is expected to be a fraction of the star bolometric luminosity. The bowshock of $\lambda$ Cep is detected by {\it WISE} and {\it IRAS} at different IR bands with  maximum values between 22 and 60\,${\mu}$m; assuming that the peak of the radiation occurs between these wavelengths, we compute the dust temperature as $T_{\rm d } \sim \frac{1}{6}hc/\lambda_{\rm max}k \sim 60$\,K. The photon distribution is considered a normalized black body at $T_{\rm d}$,  assuming that the bowshock is optically thin in the IR. The normalization warrants that the bolometric IR luminosity is $\sim$ $10^{-1}$\,$L_{*}$.

We compute the theoretical IC luminosity $L_{\rm IC, i}^{\rm t}$ at the energy $E_{\rm i}$ and fit the observed luminosity $L_{\rm \gamma, i}^{\rm o}$ detected by {\it Fermi}; we exclude from the fit the last point at $E \sim 2\times10^{10}$\,eV because of its large error bars. We use a cross-entropy method for variance minimization for performing the fit. Quantitatively, we minimize the distance $D$, which is the relative distance
 between the theoretic and observed luminosity in the norm 2:  
$D = \sqrt{\Sigma_{i}(L_{\rm IC, i}^{\rm t}/L_{\rm IC, i}^{\rm o} - 1)^2}$, where $i$ runs through all {\it Fermi} data. 

The result is presented in Figure\,\ref{sed_lam}. We obtain from this fit the injection index $\alpha \sim 2$, the maximum energy $E_{\rm max}\sim$ 110\,GeV, and $\chi \sim 0.6$ with $D = 5\times10^{-1}$. These values are in agreement with what is expected for DSA operating in the boswshock.

We can estimate the minimum magnetic field required to accelerate electrons up to $E_{\rm max}$ by equating the energy gain rate of the acceleration process and the energy loss rate (in this case dominated by wind convection). We obtain that a relatively small magnetic field $B \sim 0.06$\,${\mu}$G is required to reach $E_{\rm max}\sim $110\,GeV in this system. See the discussion for further considerations.

\section{LS 2355 as a possible  gamma-ray source} 

\begin{figure}
\centering
\includegraphics[width=8.5cm]{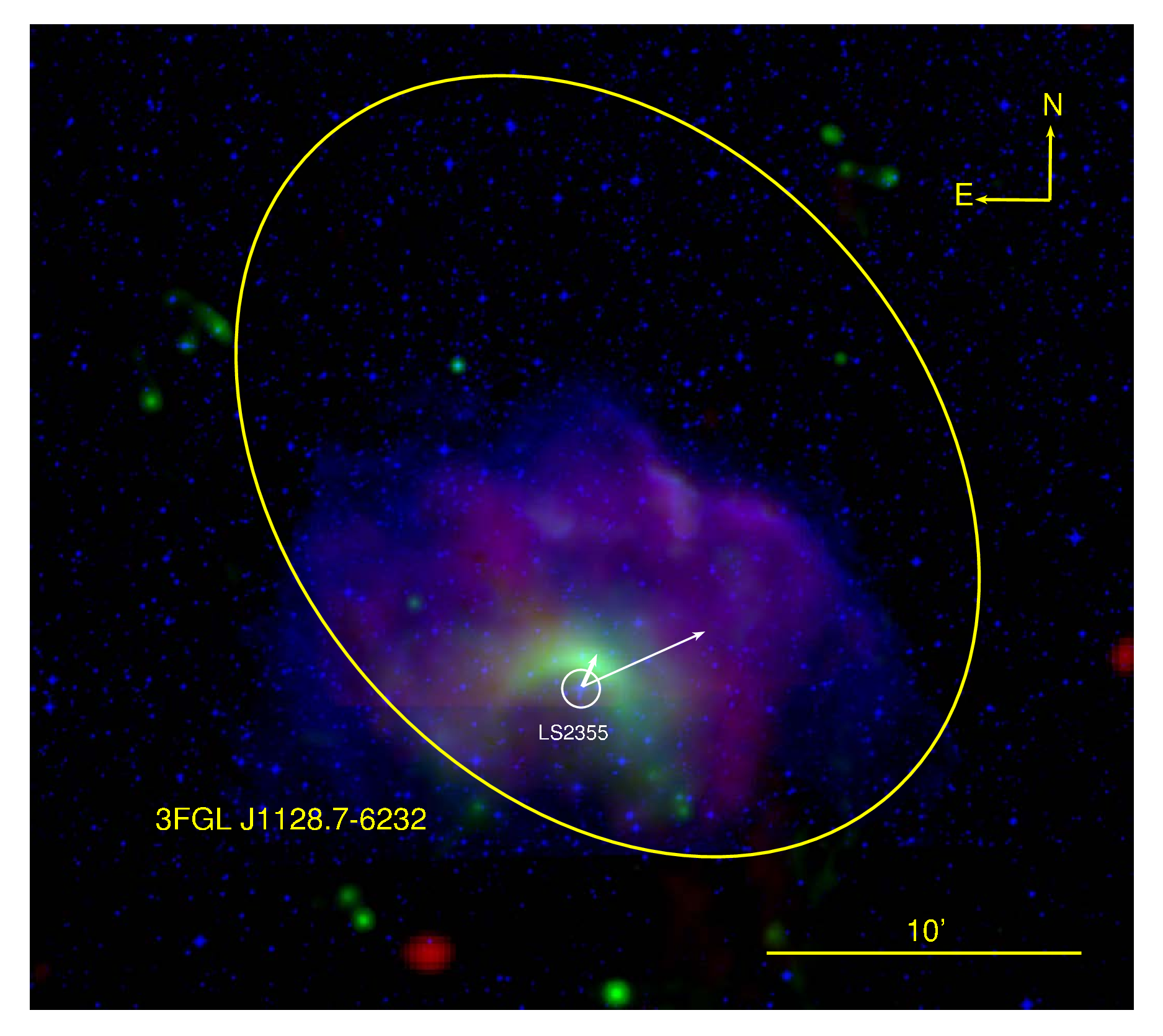}
\caption{Error ellipse at the 95\% confidence level of the unidentified gamma ray source 3FGL J1128.7-6232 projected towards the direction of the shock front created by the runaway star LS 2355. This trichromatic composition shows the SUMSS radio emission at 843 MHz in red, the {\it WISE} infrared emission at 22 $\mu$m in green and the optical emission of the DSS2 red plate in blue. The bowshock remains again as the most peculiar object of the field. The white circle marks the position of the runaway star. The white  thin and thick vectors
illustrate the observed proper motion before and after the correction for the Galactic rotation.}
 \label{elipse_LS}
 \end{figure}


LS 2355  is an early spectral type O6.5IV runaway star \citep[see][]{2014ApJS..211...10S} towards the constellation of Centaurus. 
Based on its sky location, LS 2355 has been associated with the open cluster Cru OB1  \citep[see][]{1994MNRAS.269..289K}. 
In this work, we adopt a distance of  $2.1 \pm  1.1$ kpc  derived from the parallax information furnished 
by the first data release of the \citet{2016yCat.1337....0G}. 
To better estimate proper motions, we have combined the {\it Gaia} data with other available
sources of information in order to take advantage of nearly a century-old time baseline 
(see Appendix \ref{pmdis} for details).

Figure~\ref{elipse_LS} shows the 95\% confidence error ellipse of the unidentified gamma ray source 3FGL J1128.7-6232, 
taken from  \citet{2015ApJS..218...23A},  onto a trichromatic composition. The same surveys mentioned above have been used here except for the radio layer, for which we display the Sydney University Molonglo Sky Survey (SUMSS) 
radio emission at 843 MHz in red. In this case also the bowshock appears  as the most peculiar object at all wavelengths compatible with the {\it Fermi} position ellipse. This bowshock is just at the Southern edge of  the H~II region named GAL $293.60-01.28$, whose nebulosity has been studied by several authors \citep{2000A&A...357..308G, 2009ApJ...699..469C, 2012ApJ...752..146L}.

The bowshock has an angular size of $\sim$ 111 $\times$ 74  arcsec$^2$ on the plane of the sky (see Fig.~\ref{elipse_LS}). 
The volume of the paraboloid limited by the star and the bowshock vertex is about $(3.0 \pm 0.5) \times 10^{55}$ cm$^3$.
We derive a RSR velocity of $23 \pm 5$ km s$^{-1}$ for LS 2355  whose position angle,
once corrected for Galactic rotation, clearly points towards the bowshock vertex (see  Appendix \ref{rotgal}).
The modulus of the RSR velocity found has a moderately low value for a runaway star. 
The prominent presence of the bowshock is then likely to be further enhanced by over-dense
environment as the star motion impinges into the H II region. From the image we also  estimate $R$ that gives $\sim 0.6$\,pc; in this case the velocity is almost in the plane of the sky and no angle correction is needed.

\begin{figure}
\centering
\includegraphics[width=8.5cm,angle=0]{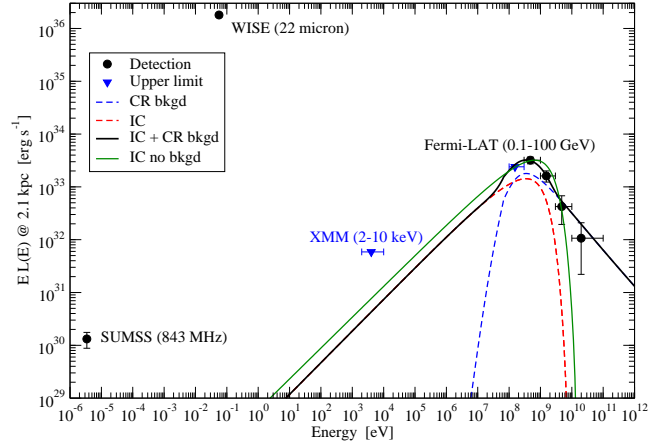}
\caption{Fitted SED for the LS 2355 bowshock at an assumed distance of 2.1 kpc. 
SUMSS (843 MHz), WISE (22 $\mu$m), X-ray upper limit (2-10 keV) and {\it Fermi}-LAT data  points are 
shown. Blue triangles indicate upper limits. The model fit curve (black line) corresponds to the sum of the background component produced by
cosmic ray protons (blue dashed line) plus the IC produced by the electrons in the bowshock (red dashed curve). It is also
shown in green the fit curve resulting from IC only (no background component).}
  \label{SED}
  \end{figure}
    

\subsection{Spectral energy distribution of the LS 2355 bowshock}

This  SED has been compiled from different sources and is plotted  in Fig. \ref{SED}.
The horizontal axis gives the logarithmically spaced photon energy,  $E$ in eV, and the vertical axis  the monochromatic luminosity times the energy, $E L(E)$ in erg s$^{-1}$.
The value of the distance quoted above has been used.
Radio and IR emission are estimated based on SUMSS and {\it WISE} images, respectively.
We also searched for diffuse X-ray emission associated with the bowshock. If present, it must be below the detection limits of the XMM-{\it Newton} archive data. Thus, a 4$\sigma$ upper limit for the X-ray flux has been estimated from  XMM-{\it Newton} archival data (Obs. ID 0672320101) by assuming 
a power-law spectrum with $\Gamma = 2$ and a hydrogen column density of $N_H \sim 2.5 \times 10^{21}$ cm$^{-2}$. This latter value is derived
from the star colour excess $E(B-V) \simeq 0.43$ mag, that translates into a visual extinction value of $A_V \simeq 5.3 \times 10^{-22} N_H \simeq 1.3$ mag.
The {\it Fermi}-LAT spectrum of 3FGL J1128.7-6232 is also included in Fig. \ref{SED}.
The numerical values of the data are given in the Appendix (Table \ref{fluxesLS}).

\subsection{Model and fitting}\label{sec:model2}

We fitted the observed emission with the bowshock model as done previously in Sect.\,\ref{sec:model}. The wind parameters of this star are not well-determined. However,  since LS 2355 shares the same spectral type as $\lambda$ Cep, we have assumed similar values for the wind and identical bolometric luminosity. For a wind mass-loss rate of $\dot{M} = 10^{-6}$\,$M_{\odot}$\,yr$^{-1}$ and wind velocity $V_{\rm w} \sim 2\times10^{3}$\,km\,s$^{-1}$, $P_{\rm w}$ is $\sim$ $10^{36}$\,erg\,s$^{-1}$. We also adopted the same IR photon field for the IC scattering.

The {\it Fermi} position ellipse of the source 3FGL J1128.7-6232 also contains the HII region. Then there is the possibility that the gamma emission detected by {\it Fermi}, or at least part of it, might be produced by the background cosmic rays (CRs) ---protons mainly--- through $pp$ inelastic collisions with the protons of the HII region. In order to investigate this possibility we estimate the $pp$ emission considering a proton cosmic-ray flux in the Galaxy $J_{\rm CR}^{\rm p}(E) = 2.2 \left(\frac{E}{\rm GeV}\right)^{-2.75}\,\,{\rm cm}^{-2}\,{\rm s}^{-1}\,{\rm sr}^{-1}\,{\rm GeV}^{-1}$ \citep[][]{1996A&A...309..917A}. The density of the region can be estimated from the interaction between the stellar wind and the ambient material. The distance $R$ can be thought as the stand-off point of the bowshock, where the ram pressure of the wind balances with that from the incoming ISM (in the star reference frame); then
\begin{equation}\label{eq:R}
R = \sqrt{ \frac{V_{\rm w}\dot{M}}{4{\pi}\,\rho_{\rm a} V_{\star}^{2}}},
\end{equation}
\noindent where $\rho_{\rm a}$ is the ambient density. Using the parameters for the wind given above and the measured value of $R$, together with an assumed mean ISM gas with $\mu = 2.3\times 10^{-24}$\,g, we  estimate $n_{\rm a} \sim 1.5 \times 10^{3}$\,cm$^{-3}$, which is reasonable for an HII region \citep[e.g.,][]{2000asqu.book.....C}. 
With this density and the proton cosmic-ray flux, assuming a cloud with the semi-major axis of the  {\it Fermi} ellipse of 8.4 \,pc, we calculate this expected background component. This component alone cannot explain the {\it Fermi} source, higher radiation levels are needed. Therefore for fitting the emission we consider the sum of this background component plus the IC produced by the electrons accelerated locally.

The result  is shown in Figure\,\ref{SED}. The fitting is done with the data points of the lowest energies (see the Appendix, Table \ref{fluxesLS}). This produces a better fit besides the error bars are
smaller at these energies. From this fit we obtain $\alpha \sim 1.8$, $E_{\rm max}\sim${90}\,GeV, and $\chi \sim 2.7$ with $D = 3\times10^{-1}$. This high value of $\chi$ may seem unrealistic. It actually might reflect our ignorance of the true parameters of the system. It could be lower if, for instance, convection of relativistic particles is not effective, something that might occur if Kelvin$-$Helmholtz instabilities affect the streams \citep[e.g.,][]{1998A&A...338..273C}. 
A more elaborated model would easily produce a better fit. The minimum magnetic field required to accelerate an electron up to {90}\,GeV is relatively small, as in the previous case: about 0.1\,${\mu}$G. With this value,  the synchrotron emission lies well below the radio data point. In any case, the SUMSS flux density contains a significant contribution of unrelated thermal emission from GAL $293.60-01.28$.

 Although it is not possible to get rid of the background component we analyze the case in which this is absent. This produces also a good fit, as can be seen in Figure 4, green line. For this fit we get $\alpha = 1.8$,
$E_{\rm max} \sim 125$\,GeV, ${\chi} \sim 5$ and $D = 6.5 \times 10^{-1}$.

\section{Discussion and conclusions}

The lack of detected radio emission from the bowshock of the stars under study does not allow us to obtain a value for the magnetic field directly from observations. Instead, we can estimate the magnetic field needed to achieve the inferred maximum energies with a diffusive shock acceleration process, taking into account that the actual magnetic field can be higher. The electrons are more likely accelerated at the reverse shock, in the wind, because this shock is faster and adiabatic. Hence, the magnetic field in the acceleration region is that of the stellar wind, which depends on the stellar magnetic field. This field might also be shock-compressed, depending on its geometry.

We can estimate the synchrotron radiation expected from the same electron population that produces the gamma rays when they interact with the ISM magnetic field of $\sim 5$\,$\mu$G, as electrons eventually escape into it. This radiation can  be considered as a lower limit.
We obtain for both  $\lambda$ Cep and LS 2355 at $E \sim 6\times 10^{-6}$\,eV (1.4\,GHz), an intrinsic radio luminosity of $L_{\rm synchr} \sim 10^{29}$\,erg\,s$^{-1}$. This value is lower than the SUMSS data for LS 2355, but it is of the same order, or higher, of the VLA radio upper limits for the sources studied in \citet{2017MNRAS.471.4452D}.

We can compare our predictions with estimates made for similar sources. Firstly, our results are in agreement with the upper limit obtained for bowshocks in the H. E. S. S. Collaboration \citep{2017arXiv170502263H}, simply because there is almost no emission at $E > 100\,$GeV. The sources studied in \citet{2017MNRAS.471.4452D} are not so powerful as those studied here. The mechanical power in the wind of $\lambda$ Cep and LS 2355 is very high, and hence there is more power available to be converted into relativistic electrons. The maximum energies fitted for 4 out of the sample of 5 by De Becker et al. are also of the order of hundreds GeV, as the ones obtained here. The gamma emission for those sources at {\it Fermi} energies is between $10^{30}$ to $10^{32}$\,erg\,s$^{-1}$. This latter value is of the order of the value of 3FLG J2210.1+5925. However, the computed  {\it Fermi} upper-levels for these sources in \citet{2014A&A...565A..95S} are almost two orders of magnitude higher.

In the case of the star BD+43$^{\circ}$3654, modelled by \citet{2010A&A...517L..10B} to fit the detected radio emission, the electrons are cooled mainly by synchrotron radiation and not by IC scattering as assumed here; the best-fitted magnetic field found by \cite{2010A&A...517L..10B} is $\sim$ 50\,$\mu$G. Although the expected gamma emission is $\sim$ $10^{32}$\,erg\,s$^{-1}$ of the same order as the estimated here for $\lambda$ Cep, the spectrum is shifted towards higher energies, peaking around 1\,TeV. 

The evidence presented so far is consistent with the bowshocks being the true emitters of {\it Fermi} gamma-rays. However, to be cautious we need to better assess the chances of alternative counterpart candidates. To do so, we also carried out an additional crossmatch between {\it Fermi} \citep{2015ApJS..218...23A}, XMM-DR6 (\citet{2016yCat.9050....0R}) and NVSS (\citet{1998AJ....115.1693C}) catalogues and we found several point-like objects with X-ray counterparts in the field covered by the {\it Fermi} error ellipses for both sources considered in this paper. 
At most, only one compact X-ray source (3XMM J221043.5+592239), which could have a very weak NVSS radio counterpart (NVSS 221043+592236) located at $\alpha$=22$^h$10$^m$43.9$^s$, $\delta$=$+$59$^{\circ}$22$^\prime$36.1$^\prime$$^\prime$; J2000, could provide an alternative identification to the \textit{Fermi} gamma-ray source 3FLG J2210.1+5925. However, the object is very faint in X-ray and radio and we could not obtain any further evidence of its possible association. We did not find any alternative counterpart candidate to the \textit{Fermi} gamma-ray source 3FGL J1128.7-6232.

In addition, the possiblility of a pure chance association for both associations presented here should be also analyzed. In order to estimate the a priori probability of random coincidence of the stars under study with a {\it Fermi} source, we carried out a series of Monte Carlo simulations following the approach outlined in \cite{1999A&A...348..868R}. With $1^{\circ}$ binning, and running up to $10^6$ {\it Fermi} synthetic populations, we found a total 4820 and 11921 coincidences at the positions of $\lambda$ Cep and LS 2355, respectively. Thus, the corresponding probabilities are estimated at the level of about 0.5\% and 1.2\%. Running the simulations with $2^{\circ}$ binning does not change significantly these numbers. Therefore, the idea of a physical association with the 3FGL sources is feasible and worth of further consideration.

Finally, we comment briefly on why there are so few gamma-ray candidates among the relatively large population of massive runaways. It is a well-known fact that isolated massive stars are not efficient at producing relativistic particles. This is likely related to the absence of strong terminal shocks in their wind. In the case of runaway stars the mass swept by the bowshock accumulates in front of the star so the wind impacts on this material and a strong reverse shock in generated with a velocity of $\sim10^3$ km s$^{-1}$.  Such velocity can result in efficient particle acceleration. Non-thermal radiation will be then produced either by synchrotron or IC losses. In case of high magnetic fields, synchrotron cooling might dominate, but then the gamma-ray counterparts will be weak or absent (as in the case of BD+43$^{\circ}$3654 ). In order to IC cooling to dominate, the magnetic field should be low, but in addition an external photon field is necessary. This field is provided by the material swept by the star, whose density depends both on the ambient conditions and the proper velocity of the star. Hence, only fast stars moving in a dense medium of low magnetization are favored. It is not surprising that just few stars satisfy such a stringent criterion. In addition, the expected luminosities are not very high, as we have seen in this work and the pervious papers by \cite{2012A&A...543A..56D,2014A&A...563A..96D}. So only nearby sources can be detected with the current sensitivities. All these conditions are fulfilled by the two candidates presented here, which seem to be exceptional.

Summarizing, in this work we propose to identify two {\it Fermi} sources with the bowshocks of the massive runaway stars $\lambda$ Cep and LS 2355. These objects are the most promising candidates lying inside the $95\%$ error ellipses of the gamma-ray data. Using a simple model we fit the observed emission for reasonable parameters, concluding that both systems are in principle capable of producing the gamma radiation detected by  {\it Fermi} and in agreement with the existing multiwavelength data. We showed that a chance association is not very likely. In the case of  3FGL J1128.7-6232 we have considered a possible additional contribution from the irradiation of a nearby HII region by ambient cosmic rays. We conclude that both bowshocks might be true gamma-ray emitters and deserve further investigation.

\acknowledgements
This work was supported by the Agencia Estatal de Investigaci\'on grant AYA2016-76012-C3-3-P from the Spanish Ministerio de Econom\'{\i}a y Competitividad (MINECO), and  the Consejer\'{\i}a de Econom\'{\i}a, Innovaci\'on, Ciencia y Empleo of Junta de Andaluc\'{\i}a under research group FQM-322, as well as FEDER funds. M.~V.~d.V. acknowledges support from the Alexander von Humboldt Foundation. G.E.R., thanks support from PIP 0338 (CONICET) and AYA 2016-76012-C3-1-P.


\bibliography{referencias_josep.bib}

\begin{table}
\caption{Astrometric catalogue positions for LS2355}              
\label{astrometLS}      
\centering                                      
\begin{tabular}{l l l l l}          
\hline\hline                        
Epoch  & ICRS Right Ascension & ICRS declination &  Errors  & Reference \\    
(year)  &               (deg)              &        (deg)           & (mas)  &    \\ 
\hline                                   
1923.446 & 172.226129145  & $-$62.652808333 &   295,   87 &  AC2000, \citet{1998yCat.1247....0U}\\
1970.4425 & 172.22592624  & $-$62.652760277 &   23,  21 & CPC2, \citet{1999yCat.1265....0Z} \\
1991.25  & 172.22579366 &    $-$62.65273602 & 0.57,  0.61 & Hipparcos, \citet{2007ASSL..350.....V}\\
1998.235 & 172.2257600 & $-$62.6527306 & 3,  3 & UCAC5, \citet{2017yCat.1340....0Z} \\
2000.0 & 172.2258018 &  $-$62.6527298  &  7.3, 7.3 & SPM 4.0, \citet{2011AJ....142...15G} \\
2015.0 &  172.2257001823 &  $-$62.6527254619 &  0.210,  0.215 & {\it Gaia} Collaboration (2016)  \\
\hline                                             
\end{tabular}

\end{table}


\begin{table}
\label{fluxesLam}      
\centering     
\caption{SED  values for $\lambda$ Cep bowshock}
\renewcommand{\arraystretch}{1.0}
\renewcommand{\tabcolsep}{0.09cm}
\tiny{
\begin{tabular}{c c c c c c c c c}          
Telescope & {\it WISE}  & {XMM-{\it Newton}}  &    \multicolumn{5}{c}{{\it Fermi}-LAT} \\    


\hline\hline  
& &&&&&&&\\
$E$ [eV]
& 5.64$_{-0.56}^{+0.56}\times 10^{-2}$ 
& 4.02$_{-2.02}^{+5.98}\times 10^{3}$
& 1.54$_{-0.54}^{+1.46}\times 10^{8}$
& 4.76$_{-1.76}^{+5.23}\times 10^{8}$
& 1.54$_{-0.54}^{+1.46}\times 10^{9}$
& 4.76 $_{-1.76}^{+5.23}\times 10^{9}$
& 1.98 $_{-0.98}^{+8.02}\times 10^{10}$ \\    

& &&&&&&&\\
$EL(E)$ [erg s$^{-1}$]
&9.39$_{-1.17}^{+1.17}\times 10^{35}$  
&3.37$\times 10^{30}$ 
&2.65$_{-0.55}^{+0.65}\times 10^{32}$    
&2.08$_{-0.27}^{+0.27}\times 10^{32}$ 
&8.91$_{-1.80}^{+1.88}\times 10^{31}$ 
&9.38$_{-9.38}^{+10.72}\times 10^{30}$ 
&3.00$_{-3.00}^{+66.03}\times 10^{29}$ \\
& &&&&&&&\\
\hline                                             
\end{tabular}
}
\end{table}


\begin{table}
\label{fluxesLS}      
\centering     
\caption{SED values for the LS 2355 bowshock}
\renewcommand{\arraystretch}{1.0}
\renewcommand{\tabcolsep}{0.09cm}
\tiny{
\begin{tabular}{c c c c c c c c c c}          
Telescope & SUMSS   & {\it WISE}  & { XMM-{\it Newton}}  &    \multicolumn{5}{c}{{\it Fermi}-LAT} \\    

\hline\hline  
& &&&&&&&\\

$E$ [eV]
& 3.44$_{-0.01}^{+0.01} \times 10^{-6}$ 
& 5.64$_{-0.56}^{+0.56}\times 10^{-2} $ 
& 4.02$_{-2.02}^{+5.98 }\times 10^{3}$
& 1.54$_{-0.54}^{+1.46 }\times 10^{8}$
& 4.76$_{-1.76}^{+5.23}\times 10^{8}$
& 1.54$_{-0.54}^{+1.46}\times 10^{9}$
& 4.76$_{-1.76}^{+5.23}\times 10^{9}$
& 1.98$_{-0.98}^{+8.02}\times 10^{10}$ \\    

& &&&&&&&\\

$EL(E)$ [erg s$^{-1}$]
&1.32$_{-0.44}^{+0.44}\times 10^{30}$ 
& 1.79$_{-0.14}^{+0.14}\times 10^{36}$  
& 5.87$\times 10^{31}$ 
& 2.40 $\times 10^{33}$    
& 3.20$_{-0.46}^{+0.46}\times 10^{33}$ 
&1.61$_{-0.38}^{+0.39}\times 10^{33}$ 
&4.24$_{-2.30}^{+2.54}\times 10^{32}$ 
&1.07$_{-0.84}^{+1.03}\times 10^{32}$ \\

& &&&&&&&\\

\hline                                             

\end{tabular}
}
\end{table}

\appendix

\section{Target proper motions and distances} \label{pmdis}

Proper motions and radial velocities are required in order to correctly understand the kinematics of runaway stars.
In the case of $\lambda$ Cep and LS2355, estimates of their heliocentric radial velocities ($-75.1 \pm 0.9$ km s$^{-1}$ and $-3.0 \pm 4.5$, respectively)
 are available from the Pulkovo compilation \citep{2006AstL...32..759G}. 
 Concerning proper motions, these parameters are harder to be
 accurately measured given the need of multi-epoch observations.

 $\lambda$ Cep is a very bright star not yet available in the {\it Gaia} DR1. Given its brightness, it also easily saturated historic photographic
 plates rendering difficult the measurement of accurate positions. Therefore, we directly adopted the Hipparcos proper motions available in the
 literature: 
$\mu_{\alpha_{ICRS}} \cos{\delta_{ICRS}} = -7.2 \pm 1.6$ mas yr$^{-1}$ 
 and  
$\mu_{\delta_{ICRS}} = -11.06 \pm 0.44$ mas yr$^{-1}$
 \citep{2007ASSL..350.....V}.
 
\begin{figure}
\plottwo{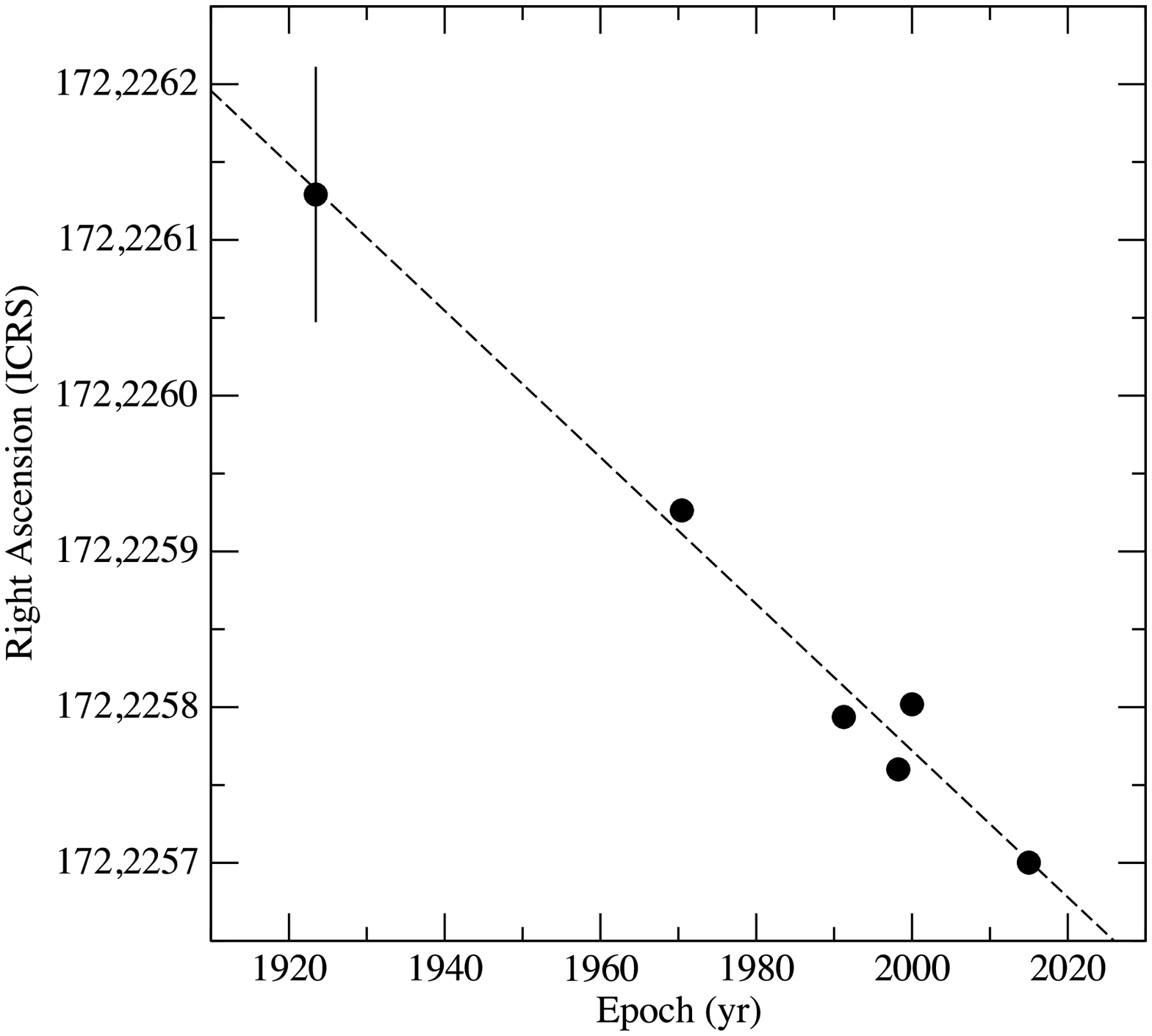}{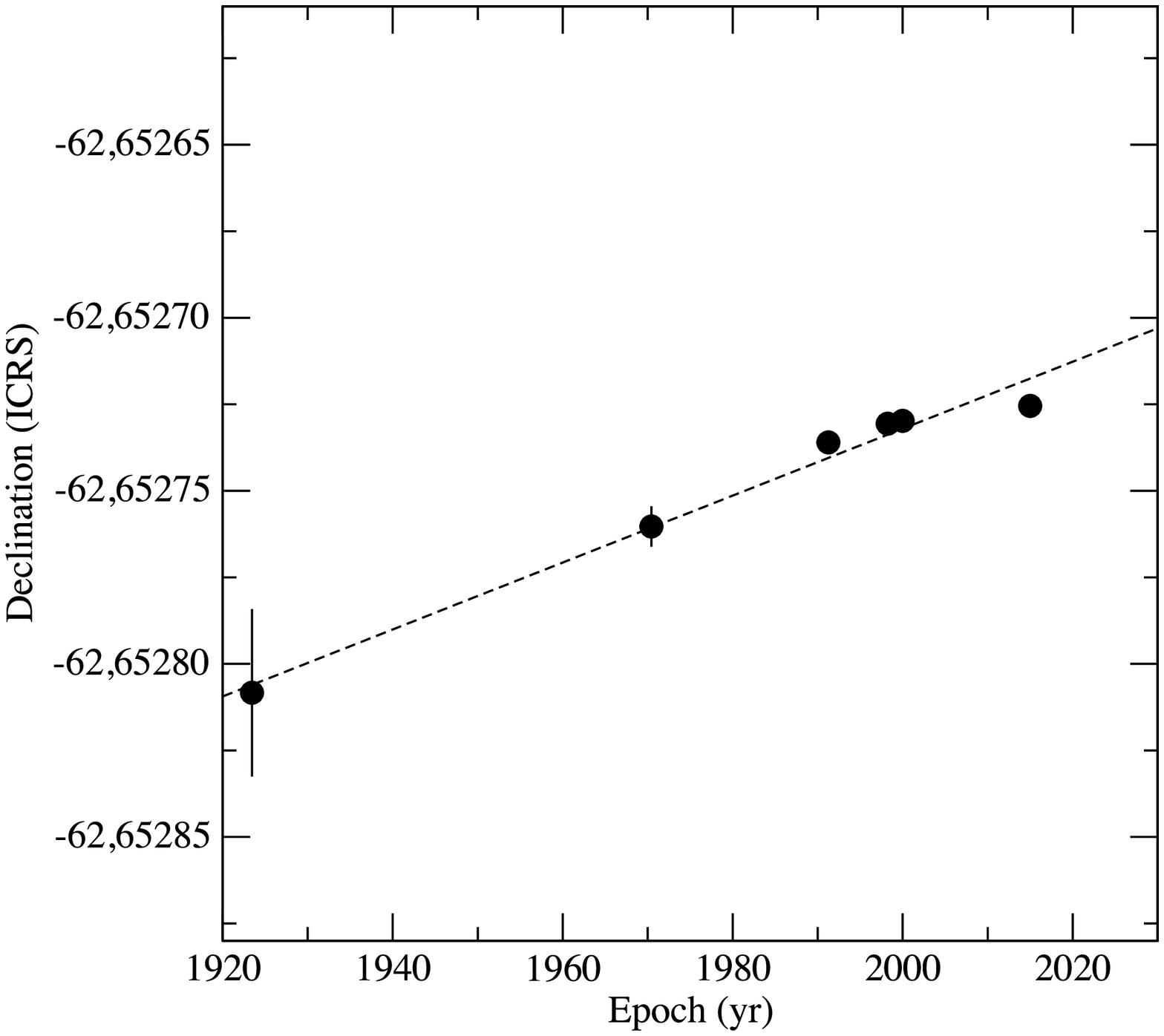}
\caption{Least squares fit to proper motion in right ascension and declination for LS2355.}
 \label{RA_pm}
 \end{figure}

With respect to the LS2355 star, its astrometric positions are available in the literature, and we have compiled
 them as shown in Table \ref{astrometLS}. When the catalogue positions were not directly expressed in the International Celestial Reference System (ICRS), we transformed them for appropriate comparison using the rotation matrices in \citet{1983A&A...128..263A} from B1950.0 to J2000.0 and/or in \citet{2004A&A...413..765H}
 from J2000.0 to ICRS. The fit results are displayed in Fig. \ref{RA_pm} yielding values of
$\mu_{\alpha_{ICRS}} \cos{\delta_{ICRS}} = -7.8 \pm 0.5$ mas yr$^{-1}$ 
and $\mu_{\delta_{ICRS}} = +3.5 \pm 0.3$ mas yr$^{-1}$ in ICRS right ascension and declinations, respectively.
The least squares fit was carried out given equal weight to all points in order to avoid an excessive bias towards the
still preliminary \citet{2016A&A...595A...1G} data. The {\it Gaia} point apparently does not follow the general trend in declination (see Fig. \ref{RA_pm}).

Concerning distances, we adopt for both stars the values corresponding to their latest trigonometric parallax available from
{\it Gaia} and Hipparcos \citep{2016A&A...595A...1G, 2007ASSL..350.....V}. The values used in the SED calculations are thus 
 $0.6\pm0.1$ and $2.1\pm 1.1$  kpc for $\lambda$ Cep and LS2355, respectively.

\section{Correcting for the rotation of the Galaxy}  \label{rotgal}

Different  proper motion vectors are displayed in the trichromatic Figs. \ref{elipse_Lam} and \ref{elipse_LS} for $\lambda$ Cep and LS2355, respectively.
The thin white arrows indicate the magnitude and direction of the proper motions as observed from a heliocentric reference frame.
We can estimate the actual velocity of the two runaway stars with respect to their RSR by combining the proper motion data, 
 the heliocentric radial velocities in the literature and the distances assumed in this paper (see Appendix A). In this process we 
 also need to adopt a plausible rotation curve of the Milky Way Galaxy and velocity  components of the Sun in its Local Standard of Rest (LSR). 
The resulting RSR velocities relative to the stellar
environments, when projected onto the celestial sphere, correspond  to the thick white arrows in Figs. \ref{elipse_Lam} and \ref{elipse_LS}.
To apply the correction for Galactic rotation, we assumed the analytic Milky Way rotation curve by \citet{1989ApJ...342..272F} and the Sun LSR velocity components
by  \citet{2010MNRAS.403.1829S}.

For $\lambda$ Cep, the position angle (PA) of the corrected proper motion measured from North to East is found to be $223^{\circ} \pm 17^{\circ}$, while the modulus of its RSR velocity amounts to $70 \pm 6$ km s$^{-1}$. Similarly, for LS2355 we obtain a PA value of $-24^{\circ} \pm 21^{\circ}$, and a RSR velocity of $23 \pm 5$ km s$^{-1}$.

In both cases it is remarkable how the PA of the corrected proper motions aligns well with the directions of the  bowshock cusps, as evidenced in the 22 $\mu$m data from WISE.  The velocity estimate for $\lambda$ Cep is certainly high, consistent with being a well known runaway star identified for a long time.
In contrast, LS2355 exhibits a not so high RSR velocity as compared to $\lambda$ Cep. However, as this object seems to be impinging a dense interstellar medium, as suggested by Fig.~\ref{elipse_LS}, the bowshock around it becomes noticeable enhanced.

\section{Numerical SED values}

The distances quoted in Appendix \ref{pmdis} (0.6 kpc for $\lambda$ Cep and 2.1 kpc for LS 2355) have been used to estimate the luminosities in the Tables 3 and 4 of this Appendix.

\end{document}